\documentclass[conference]{IEEEtran}

\usepackage{caption}
\usepackage[font=footnotesize]{subfig}
\usepackage[T1]{fontenc}
\usepackage[latin1]{inputenc}
\usepackage[cmex10]{amsmath}
\usepackage{amsmath}
\usepackage{amssymb}
\usepackage{theorem}
\usepackage{wrapfig}
\usepackage{cite}
\usepackage{url}
\usepackage{multirow}

   \usepackage{graphicx}

\title{Hierarchical Composition of Memristive Networks for Real-Time Computing}

\author{
\IEEEauthorblockN{Jens B\"{u}rger\IEEEauthorrefmark{1}, Alireza Goudarzi\IEEEauthorrefmark{2}, Darko Stefanovic\IEEEauthorrefmark{2}, Christof Teuscher\IEEEauthorrefmark{1}}
\IEEEauthorblockA{\IEEEauthorrefmark{1}Portland State University, Portland, OR, USA, Email: \{jb22, teuscher\}@pdx.edu}
\IEEEauthorblockA{\IEEEauthorrefmark{2}University of New Mexico, Albuquerque, NM, USA, Email: \{alirezag, darko\}@cs.unm.edu}
}

\begin{document}

\maketitle

\begin{abstract}

Advances in materials science have led to physical instantiations of self-assembled networks of memristive devices and demonstrations of their computational capability through reservoir computing. Reservoir computing is an approach that takes advantage of collective system dynamics for real-time computing. A dynamical system, called a reservoir,  is excited with a time-varying signal and observations of its states are used to reconstruct a desired output signal. However, such a monolithic assembly limits the computational power due to signal interdependency and the resulting correlated readouts. Here, we introduce an approach that hierarchically composes a set of interconnected memristive networks into a larger reservoir. We use signal amplification and restoration to reduce reservoir state correlation, which improves the feature extraction from the input signals. Using the same number of output signals, such a hierarchical composition of heterogeneous small networks outperforms monolithic memristive networks by at least $20\%$ on waveform generation tasks. On the NARMA-10 task, we reduce the error by up to a factor of 2 compared to homogeneous reservoirs with sigmoidal neurons, whereas single memristive networks are unable to produce the correct result. Hierarchical composition is key for solving more complex tasks with such novel nano-scale hardware.
\end{abstract}


\bstctlcite{IEEEexample:BSTcontrol}

\section{Introduction}
\label{sec_intro}

Unconventional computing architectures, exploiting intrinsic dynamics for computation, have the potential to provide platforms for ``faster, less expensive and more energy efficient computing'' than current conventional architectures \cite{Crutchfield2010}. One approach to constructing such architectures is the random assembly of some form of computational substrate. In \cite{Tour2002} an example of randomly assembled molecular switches was shown to implement logic functions and in \cite{Lawson2006} a general methodology for programming randomly assembled structures was introduced. A more detailed understanding can be obtained from the comprehensive survey on nanoelectronics architectures, and software tools by Haselman and Hauck \cite{Haselman2009}.


Within the growing field of emerging nanodevices, the memristor \cite{Chua1971, Strukov2008} is one candidate for the implementation of unconventional computing architectures. Random assembly of memristive devices into larger networks has been shown \cite{Avizienis2012, Stieg2012, Sillin2013}. The ability to use these random assemblies for tasks such as higher harmonics generation demonstrates computation based on intrinsic properties of dynamical systems, as exploited by reservoir computing \cite{Jaeger2001, Maass2002}. 


Goudarzi et al. \cite{Goudarzi2014,Goudarzi2014a} showed that the tolerance of reservoir computing to fault and variation as well as its ability to compute multiple tasks simultaneously make it a suitable choice for hardware implementations using unconventional substrates. Memristive-based reservoir computing, using random or ordered networks, was shown to be able to solve simple pattern classification problems \cite{Kulkarni2012, Burger2013}. However, further analysis has shown that the computational capabilities of such networks are not easily scalable by increasing the number of memristive devices within the network. Similarly, Sillin et al. \cite{Sillin2013} recognized the need for real-time feedback for solving more complex tasks.




A classical implementation of reservoir computing, called echo state network (ESN), typically consists of a random recurrent neural network as the reservoir \cite{Jaeger2001}. Rodan and Ti{\v n}o \cite{Rodan2011} showed that random connectivity is not essential to reservoir computing and that simplified network topologies, such as a simple-cycle-reservoir (SCR), can produce competitive results. This suggests that deterministic connectivity and modularity, which are also important digital design principles, may allow us to compose larger reservoirs out of memristive networks.



In this paper, we introduce an architecture that harnesses randomly assembled memristive networks as reservoir nodes, and relies on the ability to connect them in a deterministic manner to achieve higher system complexity. By using memristive networks for nonlinear computation we build upon the results shown in \cite{Sillin2013, Kulkarni2012, Burger2013}. Based on suitable combinations of input and reservoir weight scaling we improve wave generation performance by at least $20\%$ compared with single memristive networks. In addition, we tackle the NARMA-10 task, not solvable by single memristive networks, and achieve error rates up to twice as low as conventional reservoir implementations. Based on these results we can envision high-performance, parallel computing architectures that shift computation to the memristive networks and enable much simplified CMOS layers.

\section{Architecture}
\label{sec:architecture}

\begin{figure*}
\def \w{2in}
\centering
\subfloat[Memristor switching]{
\includegraphics[width=\w]{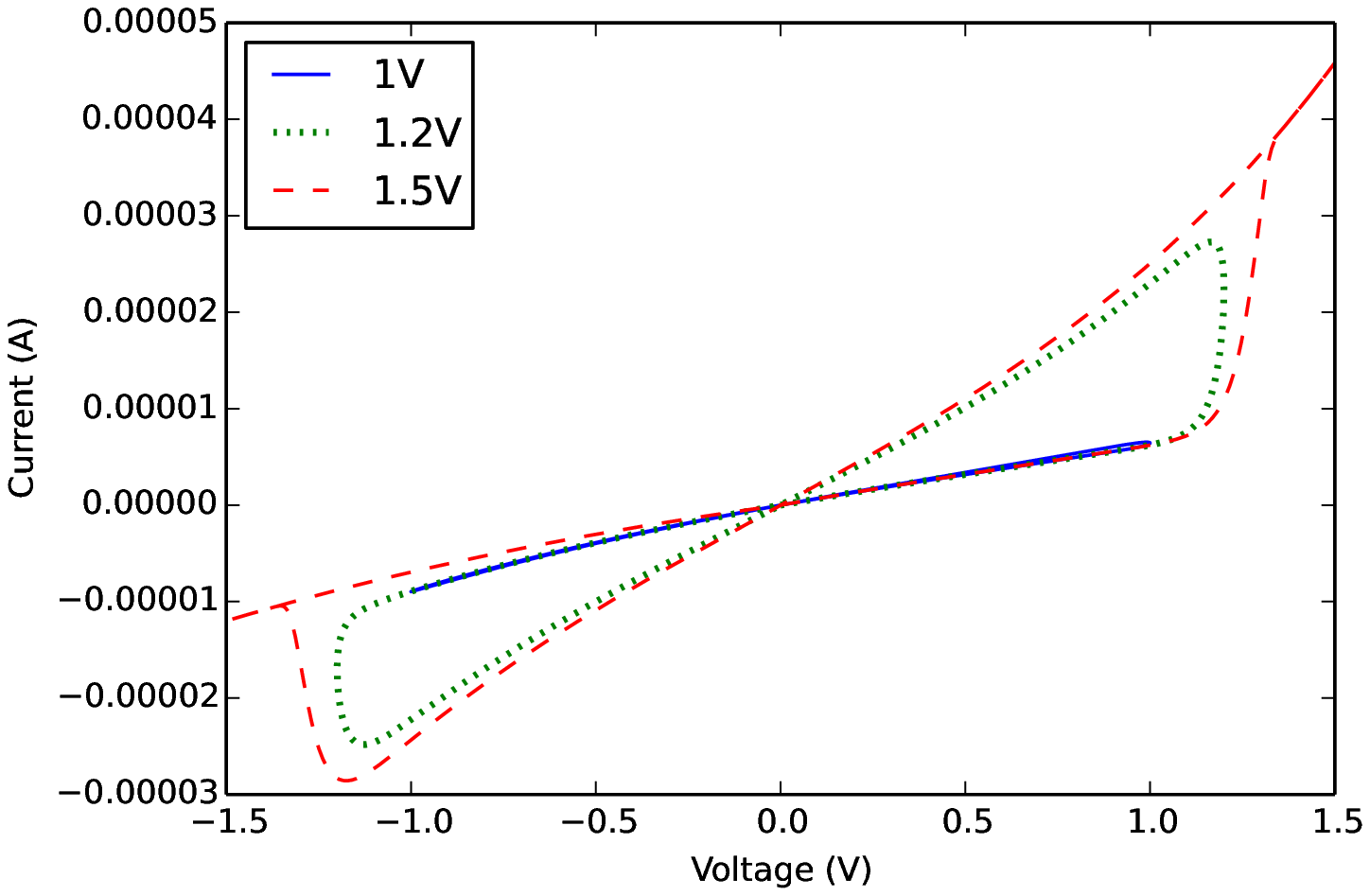}
\label{fig:memr_switching}
}
\subfloat[Memristive network]{
\includegraphics[width=\w]{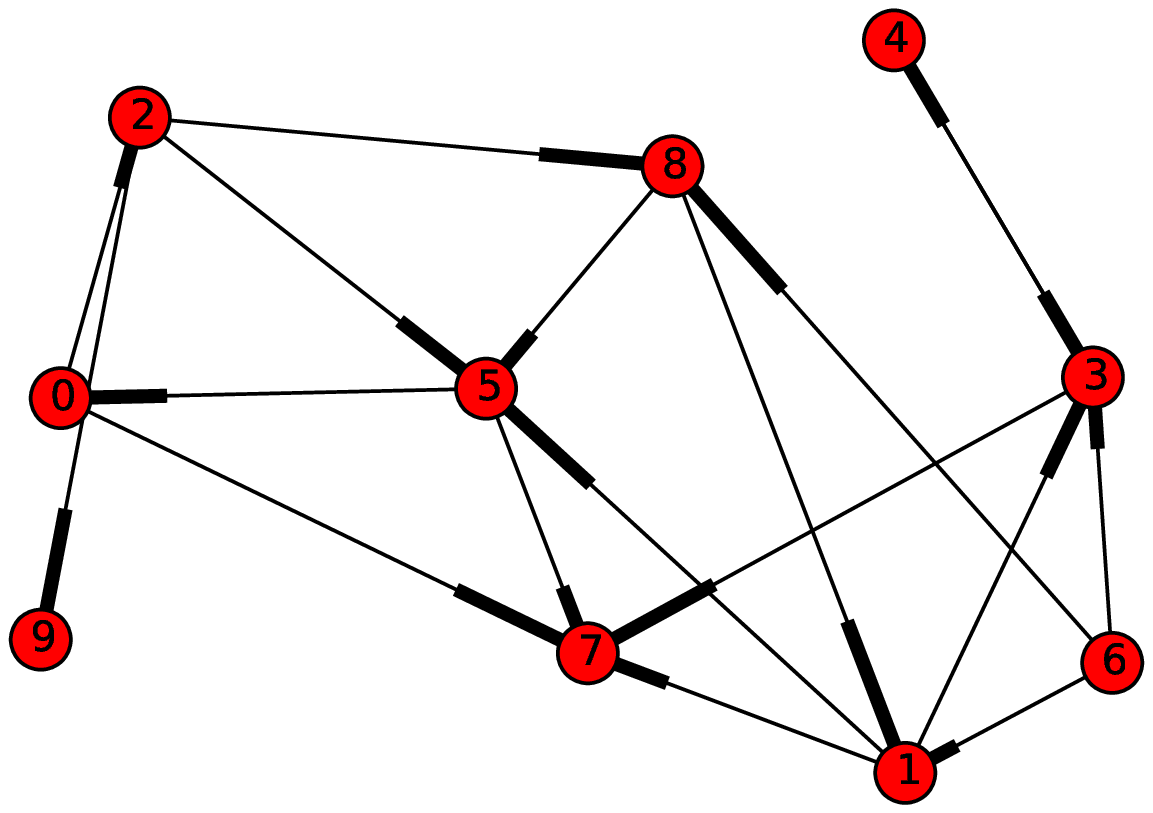}
\label{fig:memr_reservoir}
}
\subfloat[Memristive SCR]{
\includegraphics[width=\w]{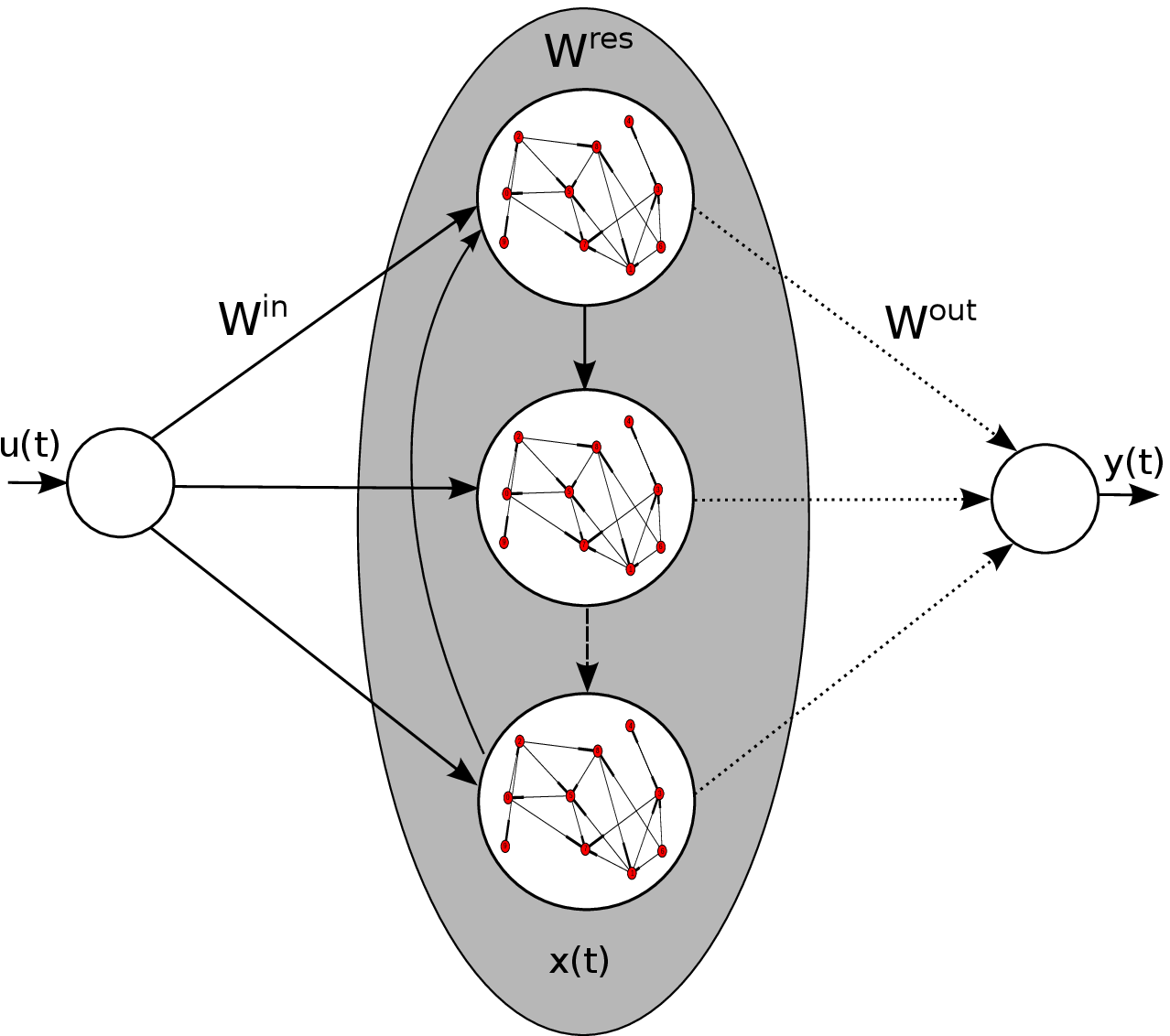}
\label{fig:memr_scr}
}
\caption{Architecture of memristive simple-cycle-reservoir (SCR). (a) Amplitude dependent memristive switching characteristics for a 10Hz applied sine wave. Low amplitudes cause no state switching as the switching threshold is not reached. Increasing the amplitude will result in non-saturated state switching. Further increasing the amplitude will result in near-binary switching due to fast saturation of the memristive state variable; (b) example of a randomly assembled memristive network. The red circles indicate nodes in which memristive devices (links between nodes) connect. The network size and device density is controlled by the number of nodes $N$ and the in-degree $K$ (number of connections per device); (c) simple cycle reservoir as in \cite{Rodan2011}. Instead of analog neurons, memristive networks provide the input-output-mapping of each SCR node.}
\label{fig:architecture}
\end{figure*}

Memristive reservoir computing relies on dynamic state changes of the memristive devices \cite{Burger2013}. For nonlinear device types voltages greater than a given threshold are required to drive these memristive state changes. Increasing the computational capacities of memristive networks faces two main challenges, both related to the network size and density, which determine the effective voltage drops over all devices within the network. Too low device voltages would not cause any memristive state changes required for computation, whereas too high voltages can cause damage to devices. Second, increased network size also raises the question of how to interface to additional network states. Due to the differences in the predicted form factors of nanodevices and of an underlying CMOS layer, an increased number of interfaced memristor signals would require spreading the memristor network over a larger area, losing some density advantages of nanodevices \cite{Likharev2005}. We will outline a modular architecture that circumvents the mentioned challenges of voltage scaling and signal interfacing by composing a larger reservoir out of small memristive networks.

The core element of our architecture is a memristive device. We rely on the device model described in \cite{Chang2013}. The current response as well as the memristive state updates are described by the following two equations.

\begin{align}
I(t) &= \left(1-w(t)\right)\sigma\left[1-\exp\left(-\beta V(t)\right)\right] \nonumber \\
    &+ w(t) \gamma \sinh\left(\delta V(t)\right) 
\label{eq:current}
\end{align}

\begin{align}
\frac{dw}{dt} &= \Lambda  \sinh\left(\eta V(t)\right) - \frac{w(t)}{\tau}
\label{eq:dw}
\end{align}


The current $I(t)$ as well as the state change $dw$ are functions of the device state $w(t)$ and the applied voltage $V(t)$. $\sigma$, $\beta$, $\gamma$, $\delta$, $\Lambda$, and $\eta$ are constants that were determined to match experimental data. $\tau$ describes the speed of the state decay of the device.

In Fig.~\ref{fig:memr_switching} we show the switching characteristics of the memristive device subject to application of a 10Hz sine wave with $1V$, $1.2V$, and $1.5V$ amplitudes. In \cite{Burger2013} it was shown how the memristive state changes affect the computation, with more continuous changes being the most beneficial. This motivates using voltages beyond the device's threshold voltage to fully exploit the resistive range. However, voltages below the threshold that do not cause memristive state changes, such as the $1V$ signal, still exhibit a nonlinear voltage-current response as described by equation \ref{eq:current}, which might be harnessed for computation. 

The next level of our architecture is the memristive network, which is randomly assembled by a set of memristive devices, represented by links (Fig.~\ref{fig:memr_reservoir}). The red circles represent nodes in which devices connect and where it is possible to interface to network states. Such memristive networks are the building blocks for composing the larger reservoir architecture.

Similar to what was shown in \cite{Likharev2005, Avizienis2012, Sillin2013} we assume an underlying CMOS grid providing vertical posts that interface to these memristive networks. This allows the composition of memristive networks into a larger reservoir by (a) applying input signals to the networks, (b) implementing interfaces to read network states (differential signal obtained from two random network nodes), and (c) signal routing and amplification between memristive networks in the CMOS layer following the ring structure with minimal connections as presented in \cite{Rodan2011} (Fig.~\ref{fig:memr_scr}). With each SCR node interfacing to a distinct memristive network (in other words a distinct nonlinear input-output function), this architecture provides the basis for heterogeneous reservoir computing. Each SCR node output is forwarded to the readout layer. A detailed description containing the readout layer is given in Section \ref{sec:experiments}.


\section{Experimental Setup}
\label{sec:experiments}

\subsection{Reservoir Computing Model}
\label{sec:rc}


In our single cycle reservoir, the nodes are ordered in a ring structure with uniform fixed weights connecting the nodes without any further adaptation \cite{Rodan2011}. Figure~\ref{fig:memr_scr} is a schematic of the SCR. The readout layer computes a linear combination of the reservoir states. The readout weights are determined using supervised learning techniques, where the network is driven by a teacher input and its output is compared with a corresponding teacher output to estimate the error; the weights can be calculated using any closed-form regression technique\cite{Jaeger2001} to minimize this error. We represent the time-dependent inputs as a column vector ${\bf u}(t)$, the reservoir state as a column vector ${\bf x}(t)$, and the output as a column vector ${\bf y}(t)$. The input connectivity is represented by the matrix ${\bf W}^{in}$ where each element is assigned the weight $v$ with signs chosen according to Bernoulli distributions. The reservoir connectivity is represented by an $N\times N$ weight matrix ${\bf W}^{res}$. In our SCR, the weights of the reservoir are uniform and equal to the magnitude of the spectral radius $|\lambda|$. Spectral radius is the largest absolute eigenvalue of the weight matrix, and determines the dynamical regime of the reservoir. For $|\lambda|>1$, the reservoir amplifies the signals over time, potentially causing chaotic dynamics, whereas for $|\lambda|<1$ the signals attenuate over time leading to contractive dynamics.

As described in Section~\ref{sec:architecture}, the state of a reservoir node is the differential between two randomly chosen signals within the memristive network comprising the node, denoted hereafter by $x_i(t)$. The time evolution of the $i^{\text{th}}$ reservoir node is given by:
\begin{equation}
{\bf x}_i(t+1) = f_i\left({\bf W}_i^{res}\cdot  {\bf x}(t) + {\bf W}_i^{in}\cdot {\bf u}(t)\right),
\end{equation}
where ${\bf W}_i^{res}$ and ${\bf W}_i^{in}$ are the $i^{\text{th}}$ row of the reservoir weight matrix and the input weight matrix respectively, and $f_i$ is the distinct transfer function of node $i$ computed by its internal memristive network (see Section~\ref{sec:architecture}). The output is generated by multiplying  an output weight matrix ${\bf W}^{out}$ of length  $N+1$ and the reservoir state vector $x(t)$ extended by a constant $1$ represented by ${\bf x}'(t)$:
\begin{equation}
{\bf y}(t) = {\bf W}^{out}\cdot {\bf x}'(t).
\label{eq:output}
\end{equation}

For training, we calculate the output weights to minimize the squared output error $E=\langle ||{\bf y}(t)-{\bf \widehat{y}}(t)||^2 \rangle$ given the target output ${\bf \widehat{y}}(t)$. Here, $||\cdot||$ is the $L_2$ norm and $\langle\cdot\rangle$  the time average. The output weights ${\bf W}^{out}$ can be calculated using any regression technique.

\subsection{Memory Capacity Task}

The linear memory capacity is a standard measure of memory in recurrent neural networks.  The memory capacity is evaluated using the capacity function $C_\phi$, which is the coefficient of determination between the output $y_t$ and the desired output $\widehat{y}_t$:
\begin{equation}
C_{\phi} = \frac{\mathrm{Cov}^2(y_t,\widehat{y}_t)}{\mathrm{Var}(y_t)\mathrm{Var}(\widehat{y}_t)},
\end{equation}
where $\phi$ is the memory length for the task. The desired output for this task is defined as:
\begin{equation}
\widehat{y}_t = u_{t-\phi}.
\end{equation}
The $\phi$-delay memory function $C_\phi$  measures how well the network can reconstruct its input from $\phi$ steps ago. Memory capacity is then calculated as a summation of the capacity function over $\phi$: $C=\sum_\phi C_\phi$. We use $1\le\phi\le10$ for our empirical estimations. In these experiments reservoirs of size $N=20$ nodes are driven with a one-dimensional input drawn from  uniform distributions on $[-0.8,0.8]$.

\subsection{Higher Harmonics Generation Tasks}
Higher harmonic generation (HHG) is a nonlinear process in which a dynamical system is excited by a signal with frequency $f$ and in turn generates signals with other frequencies not present in the input. To enable comparison with \cite{Sillin2013}, we present the following three tasks: 
\subsubsection{Sine wave generation}
For this task, the reservoir is driven with a sine wave at frequency $f$ and the output is trained to produce a sine wave at  frequency $2f$.

\subsubsection{Triangle wave generation}
For this task, the reservoir is driven with a  sine wave and the output is trained to produce a triangle wave given by: 
\begin{equation}
x(t) = \frac{8}{\pi^2} \sum_{k=1}^\infty (-1)^k\frac{\sin(2\pi(2k+1)ft)}{(2k+1)^2}
\end{equation}
\subsubsection{Square wave generation}
For this task, the reservoir is driven with a  sine wave and the output is trained to produce a square wave given by: 
\begin{equation}
x(t) = \frac{4}{\pi} \sum_{k=1}^\infty \frac{\sin(2\pi(2k-1)ft)}{(2k-1)}
\end{equation}

\subsection{Multiple Superimposed Oscillator}
Prediction of superimposed oscillators is used to test the prediction and wave generation capability of recurrent networks \cite{Koryakin201235}. To perform this test the network is usually operated in a free-running mode after training and the speed at which the output deviates from the expected wave is measured. Here we use only a restricted version of the task in which the network is trained to produce the input values $5$ ms ahead of time. The input wave form to the network is defined as: 
\begin{equation}
\widehat{y}(t) = \sin(0.2t)+\sin(0.311t) + \sin(0.42t).
\end{equation}
The coefficients are designed to make sure the attractor has a long cycle length and cannot be memorized by the network. We will use this task to test prediction ability and stability of random memristor networks.

\subsection{NARMA 10}
Nonlinear autoregressive moving average 10 (NARMA 10) is a discrete-time temporal task with 10th-order time lag. To simplify the notation we use $y_t$ to denote $y(t)$. The NARMA 10 time series is given by:
\begin{equation}
y_t = \alpha y_{t-1} + \beta y_{t-1} \sum_{i=1}^n y_{t-1} + \gamma u_{t-n} u_{t-1} + \delta
\end{equation}
\noindent where $n = 10$, $\alpha = 0.3$, $\beta = 0.05$, $\gamma = 1.5$, $\delta = 0.1$. The input $u_t$ is drawn from a uniform distribution in the interval $[0,0.5]$. This task presents a challenging problem to any computational system because of its nonlinearity and dependence on long time lags. Calculating the task is trivial if one has access to a device capable of algorithmic programming and perfect memory of both the input and the outputs of up to 10 previous time steps. This task is often used to evaluate the memory capacity and computational power of ESN and other recurrent neural networks.

\subsection{Simulation Software}

All experiments were done using the software framework OGER \cite{Verstraeten2012}, a  comprehensive reservoir computing framework that provides a variety of datasets, reservoir node types, and training methods. We augmented the existing set of reservoir nodes by a {\em memristive reservoir node} (MRN). An MRN is the above-mentioned random assembly of memristors. We compute these networks by treating them as temporarily stationary resistive networks that can be solved efficiently using the {\em modified nodal analysis} (MNA) algorithm \cite{Litovski1997}. After calculating a time step using the MNA, we update the memristive devices based on the node voltages present in the network to account for the dynamic state changes of memristors.

\section{Results}
\label{sec_results}

\begin{figure}[t]
\def \w{3in}
\centering
\subfloat[power spectral density]{
\includegraphics[width=\w]{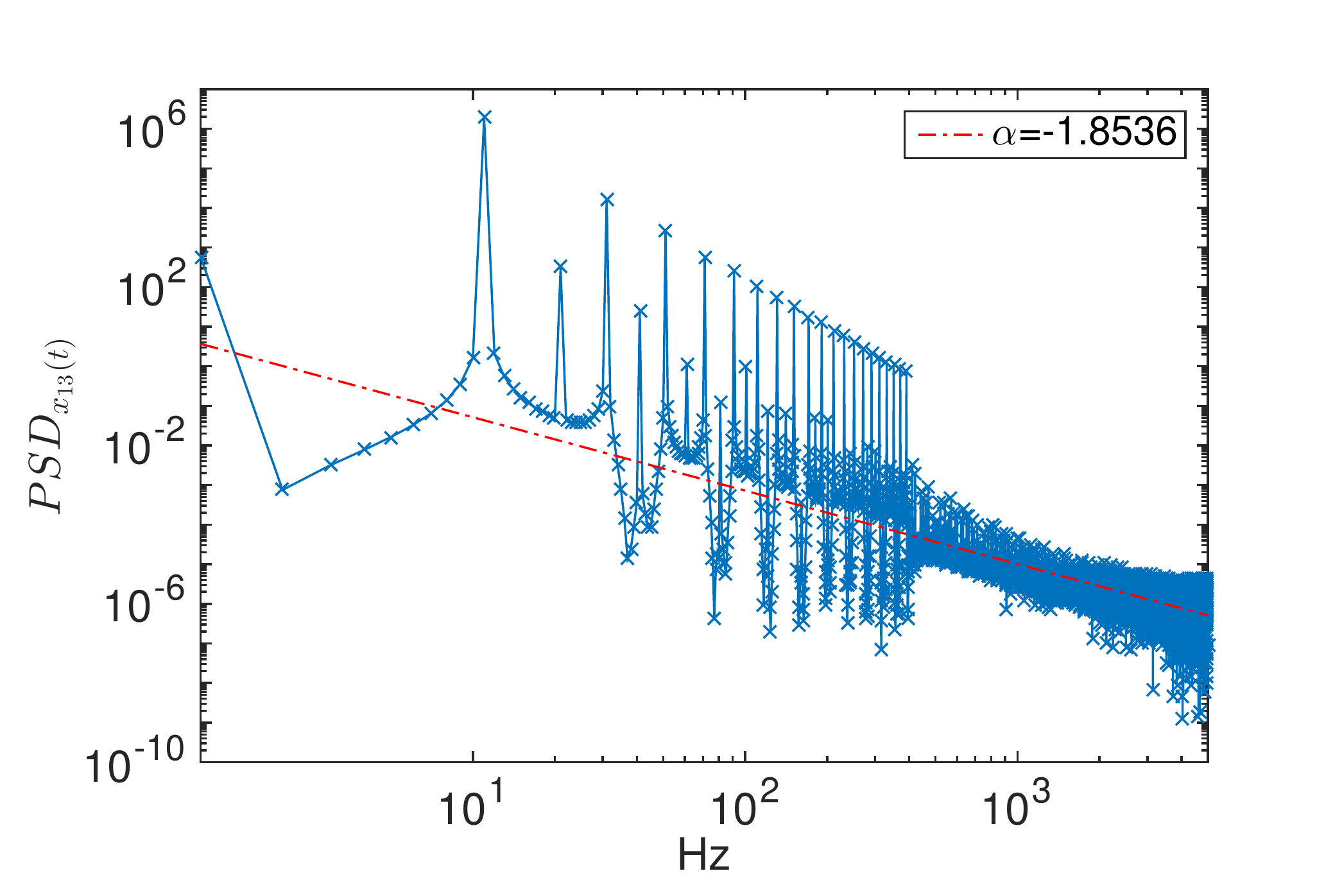}
\label{fig:examplepsd}}\\
\subfloat[memory capacity]{
\includegraphics[width=\w]{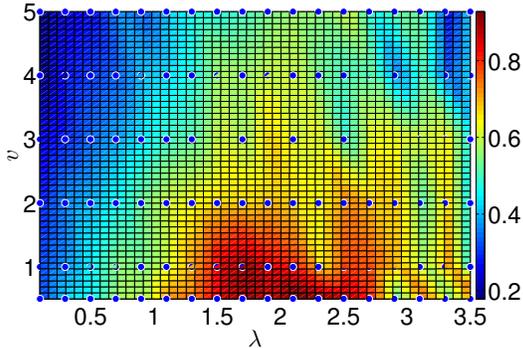}
\label{fig:memorycapacity}
}
\caption{(a) Examples of node dynamics in the reservoir (left columns) and corresponding power-spectral density (PSD) (right column). The existence of power-law PSD indicates non-trivial long-term memory at the level of single nodes similar to what was reported in \cite{Sillin2013}. (b) Memory capacity $C$ as a function of input weight coefficient $v$ and spectral radius $\lambda$. The values are normalized by system size $N=20$.}
\label{fig:memory}
\end{figure}


\subsection{Memory Capacity}
\label{sec:mc}

Figure~\ref{fig:examplepsd} shows an example of the power-spectral density of a single node in the SCR. The existence of $f^{\alpha}$ structure indicates non-trivial long-range correlation structure in the node dynamics, which suggests rich memory. A similar result was reported in \cite{Sillin2013}. Figure~\ref{fig:memorycapacity} shows the sensitivity of the memory capacity $C$ to two parameters, the input weight coefficient $v$ and the spectral radius $\lambda$, for a SCR with 20 nodes and reconstruction delay of $\phi = 10$. We observe the best memory capacity for low voltages and $\lambda$ in the range between 1.5 and 2.5. Conforming with traditional ESN larger voltages lead to more dynamic switching and nonlinear state changes, which harms the ability to preserve information. Hence we can observe the known RC trade-off between nonlinear processing and high memory capacity. In contrast to ESN, the choice of $\lambda$ underlies different considerations. In traditional ESN $|\lambda|<1$ implements a fading memory of past inputs, with longer retention for values closer to one. $|\lambda|>1$ can lead to chaotic behavior as signals circulating within the recurrent network might get amplified indefinitely. For our approach, the output of a memristive network can never be greater than the input, and is most likely to be smaller. Therefore we can allow an amplification of signals, which is indicated by the best MC for $\lambda=1.5\dots2.5$. Beyond that the system does not become chaotic, but voltage limitations cause signal saturation, which limits the memory capacity. 



\subsection{Higher Harmonic Generation} 

\begin{table}
\centering
\begin{tabular}{||c|c|c|c||}
\hline
Architecture & Task & MSE (stdev) &  $v$  \\ \hline
\multirow{3}{*}{Single network} & 20Hz sine & $0.0617 (0.0191)$ & $2$  \\
& triangle  & $0.0015 (2.45e^{-4})$ & $2$  \\
& square    & $0.0521 (0.0029)$ & $12.5$  \\ \hline
\multirow{3}{*}{SCR} & 20Hz sine & $0.0111 (0.0064)$ & $2$  \\
& triangle  & $8.14e^{-4} (1.62e^{-4})$ & $2$  \\
& square    & $0.0307 (0.0037)$ & $12.5$  \\ \hline
\end{tabular}
\caption{The best observed MSE for HHG tasks and corresponding parameters for systems passing 16 values to the readout layer.}
\label{tbl:wavegenresults}
\end{table}

\begin{figure}
\def \w{3in}
\centering
\subfloat[MSE as function of input bias]{
\includegraphics[width=\w]{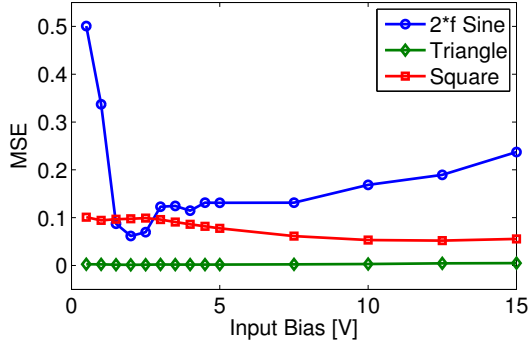}
\label{fig:wavegen_singleNetwork}} \\
\subfloat[MSE as function of readout signals]{
\includegraphics[width=\w]{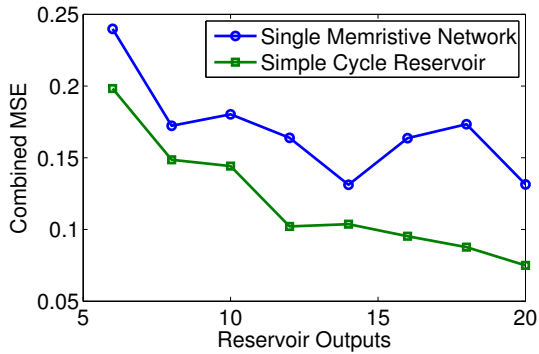}
\label{fig:wavegen_single_vs_scr}
}
\caption{(a) Wave generation MSE  single memristive network with 16 readout nodes. The curves for the individual signals are in shape similar to the ones published in \cite{Sillin2013}, even though the scale is different. (b) Comparison of cumulative MSE of single memristive network with single cycle reservoir of memristive networks as a function of readout signals. The SCR, due to the physical separation of the memristive networks can produce less dependent signals and achieve better performance.}
\label{fig:wavegen_mse}
\end{figure}

We compared single memristive networks with the memristive SCR architecture. Both setups are compared based on the number of signals extracted from them and forwarded to the readout layer. Table \ref{tbl:wavegenresults} shows the best results from setups comparable to \cite{Sillin2013}. The single memristive network had $16$ output nodes and contained $120$ memristive devices. The SCR was made up of 16 nodes with each node providing one signal and utilizing memristive networks of approximately 50 devices. Figure~\ref{fig:wavegen_singleNetwork} shows a more detailed representation of the performance of a single memristive network as a function of increasing input biases. Similar to \cite{Sillin2013} we observe the best generation of the sine wave for $v=2V$ and the square signal for larger voltages. Besides some similarities we can also observe clear discrepancies, such as the absolute MSE values and the minimal error for generating the triangle wave. Without having absolute certainty, we suspect differences in the memristive devices, constraints on the network topology, and the application/reading of input/output signals to be likely causes of these differences. 

Figure~\ref{fig:wavegen_single_vs_scr} compares the combined MSE (sine, triangle, square) of single memristive networks and of the memristive SCR. The x-axis defines the number of signals read from the corresponding architecture. For the SCR this is the same as the number of nodes. We can observe that the signals read from the SCR allow  better generation of the target signals. Due to the physical separation of the SCR nodes, the signals are less correlated and provide more features compared with signals all read from the same memristive network.

\subsection{Prediction of Superimposed Oscillators}
Superimposed oscillators are used in the recurrent neural network community to test the signal generation and prediction capabilities of a network. Here, we use the three superimposed oscillators task to demonstrate the prediction capability of our reservoir consisting of $N=20$ SCR nodes, each of which includes $55$ memristors on average. 
The sine wave is fed to the network as explained before and the output is trained to produce the correct values $5$ ms ahead of time. We repeat the experiment  with different combinations of input weight coefficient $v=\{1, 2,3,\dots, 15\}$ and spectral radius $\lambda=\{0.1, 0.3, 0.5, \dots, 2.5\}$. All results are averaged over 50 experiments for each parameter combination. Note that in this particular experiment there is no difference between the training and testing errors since the input is always fixed. We present the results using the normalized-root-mean-squared-error (NRMSE):
\begin{equation}
NMRSE=\sqrt{\frac{\langle ||{\bf y}(t)-{\bf \widehat{y}}(t)||^2 \rangle}{\langle {\bf \widehat{y}}(t)^2\rangle }}.
\end{equation}
The best average result was $NRMSE=0.17$ with standard deviation $0.07$ for $v=2$ and $\lambda=0.5$, while the best individual result was as low as $NRMSE=0.04$ for $v=1$ and $\lambda=0.7$, which is comparable to a classical ESN solving the same task.



\subsection{NARMA-10}

The NARMA-10 task, due to the need of memory, poses a difficulty that single memristive networks, as presented here, are not capable of dealing with. Experiments to verify this were done for different single memristive network sizes (75 to 350 devices) with best resulting NRMSE values of around 10 indicating this inability. 

\begin{figure}
\def \w{3.5in}
\centering
\includegraphics[width=\w]{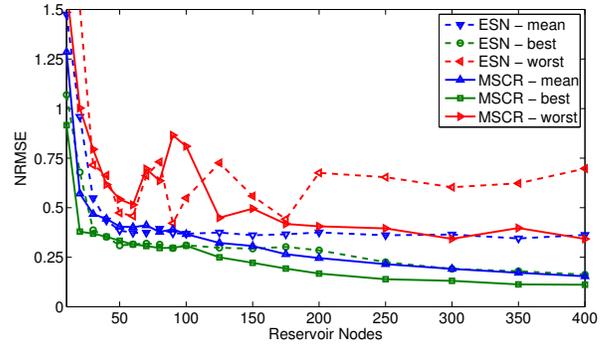}
\caption{NARMA performance for increasing SCR sizes. The dashed lines are obtained with regular sigmoidal neurons (ESN) and $v=0.01$ and $\lambda=0.75$. The memristive SCR (MSCR) data was obtained based on $v=0.5$. $\lambda=1.7$, and $52$ memristive devices on average per node.}
\label{fig:narma_perf}
\end{figure}

The memory capacity results in section~\ref{sec:mc} have shown that memory is best preserved at low voltage ranges, which also means that the memristive devices change only minimally. We found the optimal values for the MC task also leading to the best results on the NARMA-10 task. Figure~\ref{fig:narma_perf} compares the results for our memristive SCR and a regular sigmoidal neuron SCR. The low input signal range implies that the nodes of both implementations behave mostly linearly. However, due to a spectral radius greater than one and the resulting signal amplification, memristive networks experience some low-frequency internal device state changes (at a lower rate than the input signal), which adds some nonlinear processing to the reservoir. As a result of this, with a growing number of nodes, the memristive SCR continues to improve while the sigmoidal SCR plateaus at around a size of 100 nodes. We attribute this continuous performance improvement for increasing reservoir sizes to the heterogeneity of the input-output mappings (activation functions) of the memristive networks, the low frequency memristive state changes, and the resulting diverse signals used by the readout layer.


\section{Discussion}

We compared our simulation results of memristive networks to the physical realization of ``atomic switch networks'' used for reservoir computing \cite{Avizienis2012, Sillin2013}. The differences in these results highlight an important aspect of unconventional computing, namely  the high variability in structure and behavior of such computational substrates. Our results emphasize that reservoir computing allows for the utilization of varying substrates to achieve computation, despite these differences.








Initial investigations of the input-output mappings of random memristive networks have shown a wide range of behavior based on where the CMOS layer connects to the memristive networks. In extreme cases, due to the random structure, a subset of memristive networks does not contribute to computation at all. Deeper understanding of these differences and their contribution to the overall computation is key to better utilization of the hardware resources.




\section{Conclusion}

In this work we have introduced a hierarchical memristor-based reservoir computing approach. We showed that for the higher harmonics generation tasks, system parameters causing frequent memristive state switching gave the best results, outperforming single memristive networks by at least $20\%$. The NARMA-10 task, which requires memory of the past 20 inputs, performs best for system parameters that cause only sporadic memristive switching, hence better preserving the memory implemented by the recurrent structure. The variety of memristive networks allowed a better utilization of the resulting reservoir states by the readout layer, leading to better performance compared with sigmoidal neuron based reservoirs. In this work, we combined the computational capacities of memristive networks with scalability advantages of CMOS to compose complex and computationally powerful reservoir systems based on emerging nanoscale devices.


\section*{Acknowledgment}
This work was supported by the National Science Foundation under award \# 1028378, \# 1028238, and \# 1318833, and by DARPA under award \# HR0011-13-2-0015. The views expressed are those of the author(s) and do not reflect the official policy or position of the Department of Defense or the U.S. Government. Approved for Public Release, Distribution Unlimited.

\ifCLASSOPTIONcaptionsoff
  \newpage
\fi

\bibliographystyle{IEEEtran}
\bibliography{bibliography}

\end{document}